# Chain-oxygen ordering in twin-free YBa$_2$Cu$_3$O$_{7-\delta}$ single crystals driven by 20 keV electron irradiation

H. W. Seo[1], Q. Y. Chen[1], M. N. Iliev[1], N. Kolev[1], U. Welp[2], C. Wang[1], Tom H. Johansen[1,3], and Wei-Kan Chu[1]

[1]*Department of Physics and Texas Center for Superconductivity & Advanced Materials, University of Houston, Houston, Texas.*

[2]*Argonne National Laboratory, Argonne, Illinois*

[3]*Department of Physics, University of Oslo, P.O. Box 1048, Blindern, 0316 Oslo, Norway.*

We have examined the effects of 20 keV electron irradiation on [-Cu(1)-O(1)-]$_n$ chain oxygen arrangements in oxygen deficient but otherwise twin-free YBa$_2$Cu$_3$O$_{7-\delta}$ single crystals. Comparison of polarized Raman spectra of non-irradiated and irradiated areas provides evidence that electron bombardments instigate the collective hopping of oxygen atoms either from an interstitial at O(5) site to a vacant O(1) chain site or by reshuffling the chain segments to extend the average length of chains without changing the overall oxygen content. This oxygen ordering effect, while counter-intuitive, is analogous to that found in the photoexcitation induced ordering in which temporal charge imbalance from electron-hole pair creation by inelastic scattering of incident electrons causes a local lattice distortion which brings on the atomic rearrangements.

Recent observations of photoexcitation induced superconductivity and persistent normal state photoconductivity in YBa$_2$Cu$_3$O$_{7-\delta}$ (YBCO) have revealed an intriguing phenomenon involving the ordering of [-Cu(1)O(1)-]$_n$ chain-oxygen atoms [1-3]. Such ordering yields longer chains with larger *n* and hence higher carrier concentration on the Cu(2)O(3)$_2$ planes because longer chain fragments are more effective in hole doping, therefore resulting in higher T$_C$ for the same overall oxygen stoichiometry [1-4]. While the detailed ordering mechanism is still under debate, it is believed [1] that local photoexcitation of electronic states in an atom by photon irradiation leads to a fluctuation of thermodynamic potential in the nearby regions, thus providing a reduced activation energy barrier for atomic migration. Theoretical investigations made on the thermodynamics of oxygen ordering in YBCO, based on a 2D Ising model and its variants [3,5], has shown that, indeed, ordered super-patterns of non-segmented CuO chains on the CuO chain planes are energetically more favourable. This was corroborated by the observation of cell-doubled (O-II) or cell-tripled (O-III) phases in both oxygen deficient YBCO bulk crystals and thin films using neutron scattering, x-ray and electron diffraction, and electrical transport measurement [6-8].

The incident photon energy and energy transfer involved in the above-mentioned photoexcitation is ~2-3 eV [1-4, 9]. This compels us to ponder whether an electron beam that is able to impart as much energy to an oxygen atom would bring about the same ordering effect. Assuming the energy transfer comes directly from knock-on impacts, the required incident electron energy would then be 10-30 keV, which is readily available on commercial scanning electron microscopes. Electron irradiation, however, has been widely known to create crystal defects, as is the case for ~100 keV-MeV electron bombardments on YBCO superconductors [10, 11]. Therefore, there is certainly a need for verification.

Our experiments hinge on comparing the micro-Raman spectra of irradiated and unirradiated areas. The twin-free YBCO crystals used here were prepared by flux-growth technique, followed by slow cooling to room temperature to minimize residual strain, the existence of which would have complicated the Raman line shapes and peak positions. The crystal selected for the detailed measurements was a ~0.05 mm thick with lateral dimensions of $1 \times 0.7$ mm$^2$. A micrograph of the crystal taken under a polarized optical microscope is shown in Fig. 1(a), where several large area twin-free domains in the sample are obviously seen. The circle in the figure indicates to the electron irradiated area located within one of the untwinned regions. The oxygen deficiency of the crystal is estimated to be $\delta \approx 0.1-0.2$ and, furthermore, believed to be highly uniform, as judged from the magneto-optical imaging at the start of Meissner magnetic field exclusion while cooling through $T_C$ in the presence of a 100 Oe applied field (see Fig. 1 (b)).

A field emission scanning electron microscope (FESEM) was used to provide the 20 keV electron beam at a current level of ~0.2 nA. The beam was focused into a nominal spot size of 70 nm in diameter in perpendicular to the crystal plate. The beam current density was $\sim 3 \times 10^{19}$ electrons/ s cm$^2$ for a $10 \times 10$ μm$^2$ area of irradiation, amounting to a dosage of $3.6 \times 10^{18}$/cm$^2$ after 30 seconds of exposure.

The micro-Raman spectra were measured using the 632.8 nm excitation of a He-Ne laser with *aa* and *bb* scattering configurations, where *aa* and *bb* denote that the incident and scattered polarizations are both parallel to either the *a*- or to the *b*-axis, respectively. We used a laser power of 0.5 mW together with an integration time of 45 min. The low laser power was adopted to avoid possible complications from competing photoinduced ordering effects during the long integration time. The Raman data represent the top 60 nm layer, roughly the skin-depth, of the sample, where the effect of the electron irradiation is expected to be uniform.

In what follows, we will take steps to interpret the Raman data taken from the electron irradiation experiments as showing no noticeable change of oxygen content, but an obvious ordering, or healing, effect via defragmentation of the CuO chains. The healing process occurs through the irradiation-instigated collective transfer of oxygen atoms either from occupied O(5) interstitial sites to O(1) vacancy sites or by rearrangements of the [-CuO-]$_n$ chain segments such that their average length $n$ ultimately increases at sacrifice of the total number. Fig. 2 illustrates these possible routes of ordering processes.

Figure 3 shows the *aa* and *bb* polarized spectra measured under the same experimental conditions at room temperature on areas that are electron-irradiated and un-irradiated. The distinctive spectral profiles of the *aa* and *b*b spectra allow for an unambiguous identification of the *a* and *b* directions [12, 13]. At microscopic level, the structure of an oxygen-deficient YBCO is characterized by the coexistence of submicrodomains (or "submicrophases") of different oxygen arrangements (Ortho-I, Ortho-II, T, T') [14]. The Raman spectrum therefore is a superposition of spectra of these coexisting phases, the relative weight of each phase being determined by its abundance and Raman cross section, which strongly varies with the excitation laser wavelength and scattering configuration [14-18]. The Raman peak at 140 cm$^{-1}$, representing the modes of Cu(2) vibrations along the *c*-axis, is dominated by contributions from the tetragonal T ($\delta = 1$) or T' phases ($0 < \delta < 1$), partial oxygen disorder, and the position of O(2)-O(3) out-of-phase peak (337 cm$^{-1}$) is indicative of the contributions from Ortho-I submicrodomains, whereas the position at 453 cm$^{-1}$ of the O(2)-O(3) in-phase mode is suggestive of the Ortho-II or T phase [16]. The presence of Ortho-II phase ($\delta = 0.5$) is also supported by the Ortho-II apex oxygen (O4) peak at 489 cm$^{-1}$[15, 16].

The narrow peak at 228 cm$^{-1}$ and the broad band centered at ~560 cm$^{-1}$ are observed only in the bb spectra and cannot be assigned to any of the Raman allowed modes of the Ortho-I ($\delta = 0$), Ortho-II, or T submicrophases. Wake et al. [9] has studied the 228 cm$^{-1}$ peak for untwinned YBCO crystals in which it was established that the peak appears strictly when the light wave is bb-polarized and has a sharp resonance for laser energies near 2.2 eV (ours being 1.96 eV). On the basis of their laser annealing experiments, carried out using a 2.18 eV excitation, Iliev et al [16] concluded that this peak is related to the vibrations of chain-end atoms and its intensity reflects the number of chain fragments rather than the overall oxygen content. The diminution of 228 cm$^{-1}$ peak after electron bombardment thus is a manifestation of dramatically reduced number of chain ends. Given constant oxygen content, this then translates into the elongation of chain fragments. The relative weight of the chain ends would hence have to decrease, as is indeed the case, and corroborated by the decrease in intensity of the 140 cm$^{-1}$ peak assigned to that phase.

A reasonable question then is why the 20 keV electron irradiation would relocate oxygen atoms and thus enhance ordering. Since all processes took place at room temperature, thermally activated kinetics alone cannot give a full account for this phenomenon. To help our analysis, we hence borrowed supporting arguments from the observations of persistent normal state photoconductitity and enhanced $T_C$ in YBCO superconductors, both taking root in the photoexcitation induced oxygen ordering.

To begin with, we need to understand the mechanism of electronic scattering in solids. Consider a two-particle system in relativistic collision, in our case electron vs. oxygen atom, the maximal energy transferred to the host atom will be $E_{atom\_max} = 2M(E+2m_ec^2)E/[(m+M)^2c^2+2ME]$, where E is the incident electron energy, M is the host atom mass, $m_e$ is the mass of electron, and c is the speed of light [19]. The maximum energy would be transferred with a backscattering angle $\phi = 180º$, which amounts to 2.79 eV for oxygen atoms, but less for other heavier atoms such as Cu, Y, and Ba from Rutherford scattering of the incident 20 keV electron. The 2.79 eV is about what it takes to break a CuO bond (enthalpy of bond formation $\Delta H_{CuO} = 269.0 \pm 20.9$ kJ mol$^{-1}$ J.A. Kerr in [20].) In literature, Cui et al [21] assessed the energy to dislodge an oxygen atom from its lattice location to be about 1.4 eV. This being the case, then atomistic oxygen migrations should occur if scattering cross section is not taken into consideration. Nonetheless, considering all possible complex forms of inelastic scattering, starting from those of valence electrons to those of inner-shell electrons, the cross-sections of scattering go as $\sigma_{valence} > \sigma_{inner-shell} \gg \sigma_{Rutherford}$ ($\phi = 180°$) [22]. We therefore assert that most of the energy transferred would go into ionizing the constituent atoms, rather than breaking the chemical bonds.

The bonding or valence electrons thus play a major role in the ensuing atomic rearrangements in the lattice. Inelastic scattering through core-level or valence electron excitations, for instance, may cause a temporal charge imbalance in the YBCO lattice to bring on the oxygen atomic rearrangements. In essence, a shift of electrochemical potential has occurred in the sample upon the electron beam exposure, an effect similar in spirit to the photoexcitation induced charge ordering [3]. It is argued [1, 23] that electron-hole pairs are created by photoexcitation in the $CuO_2$ planes, the electrons are then localizes at oxygen vacancies, viz. empty O(1) sites in an ideal lattice, on the nearby partially oxidized Cu(1)O(1) chains. This would tend to locally enhance the orthorhombic distortion that eventually trigger oxygen atoms to hop collectively from O(5) sites to neighbouring O(1) sites, or from one O(1) site to another, as mentioned earlier, thus consummating the oxygen ordering process. The same arguments stand, we reckon, when photons are substituted with electrons. Note that though the scattering cross-section from an incident electron's collision with a free electron, $\sigma_{elastic}$, is larger than $\sigma_{valence}$, its effect on the later oxygen redistribution may be negligibly small for the less roles it plays in chemical bonding [24]. In addition, the heating effect by electron beam from phonon excitation, $\Delta T$, estimated to be ~ 27 K, is also considered insignificant [25] in the present work.

Use of electron excitations to induce oxygen ordering is an especially enticing approach as the state-of-the-art electron beams can be narrowed to a spot-size of ~1-10 nm and nano-patterning is thus possible in an electron nanoscopic setting. While a more careful study is needed as to whether electron irradiation can really lead to atomistic displacements at 10-30 keV, the mechanism of photoexcitation induced ordering may well be taken as a reasonable hint to the contrary, as we have taken here to support our assertion.

In disordered states, oxygen vacancy concentration is directly related to the reduction in $T_C$ comparing with a fully oxygenated sample, at least on the underdoped side of the phase diagram. At first glance, this is understandable as more vacancies mean more broken chains. Indeed, Tolpygo et al.[11] found that electron irradiation led to oxygen disordering at 20 keV for $\delta = 0.1$ and 0.4, representing examples of optimally doped and underdoped YBCO samples, all pointed to a similar effect of oxygen disordering shown by $T_C$ reduction and electrical resistivity increases. It is worth noting, however, that the work was carried out on thin film samples with a 3-50 nA beam current and their dosages ranging from ~$10^{19}$ to ~$10^{21}$ electrons /cm$^2$, namely 15-250 times higher in beam current and $10^3$-$10^5$ higher in dosages as compared to our irradiation conditions. We regret that, in their work, no observation was made in the dose limit as low as attempted in this work [11]. But in any event, at their high dose extremes, the knock-out effect may not be negligible anymore and its impact on the $T_C$ and normal state resistivity could hence have overshadowed the concurrent ordering effect. Meanwhile, it is worth noting, our Raman results were reproducible over long time room temperature ageing, though it is not clear as to why our sample demonstrated this extra resilience in comparison with earlier reports [1-4, 26], considering the commonly accepted high mobility of chain oxygen atoms.

In conclusion, we have studied [-Cu(1)-O(1)-]$_n$ chain oxygen ordering effect by irradiation of a twin-free YBa$_2$Cu$_3$O$_{7-\delta}$ single crystal with $\delta$ ~0.1-0.2 using 20 keV low energy electron beams. Oxygen deficient but twin-free YBCO ($0 < \delta < 1$) samples may be a mixture of submicrodomains of structural phases of different degree of oxygen deficiency averaged out to the observed $\delta$, each with its different arrangements of the O(1) and O(5) sites and chain fragments in accordance with the local stoichiometry. Based on this notion, then nominally twin-free YBCO crystals of a specific oxygen content may actually contains nano or sub-micron scale short range order or disorder. Electron irradiation triggers an ordering process to commence. In our case, the resulted ordered phase appears to survived prolonged room temperature experimental environment without showing relapse of disordering. This opens up the possibilities of tailoring oxygen order-disorder transition, i.e. manipulating the segmentation of chain oxygen in a controllable fashion to acquire the desired T$_C$, especially by use of commercially available scanning electron microscopes.

The authors are thankful to J. K. Meen and R. P. Sharma for useful discussions and technical assistance. This work was supported by the State of Texas through the Texas Center for Superconductivity and Advanced Materials at the University of Houston. Partial support by the Welch Foundation is also acknowledged.

qchen@uh.edu

**Fig 1:**

Large twin-free YBCO domains as seen under polarized microscope. The sample edges are either along a or b axis. The upper and right corners are heavily twinned. The twin boundaries along [110] direction are all clearly seen. Circled area indicates the area of electron irradiation.

Magneto-optic image (MOI) of the sample described in (a). The zig-zag pattern stems from the magnetic domains of the magneto-optic crystal used to facilitate the MOI. Beside the zig-zag, the MOI shows the uniform Meissner magnetic field repulsion at 86K. Note a small portion of the sample (Lower left) was chipped off.

**Fig 2:** Schematics of an oxygen-deficient twin-free YBCO basal plane (partially T' phase), showing i) the definition of a- and b-axes of a perfect lattice (Bottom right), and ii) various locations of oxygen atoms, vacancies, interstitials and their motions to effect the ordering.  Arrows correspond to the movements of oxygen.

**Fig 3:**  The polarized Raman spectra show twin-free YBCO crystal in the irradiated and non-irradiated areas. The position of the 140 cm$^{-1}$ peak corresponds to the T or T' phases, 337 cm$^{-1}$ to Ortho-I, 489 cm$^{-1}$ to Ortho-II.  The peak at 228 cm$^{-1}$ represents the vibration of Cu atoms at the chain ends, the number of which is measured by its intensities.  The diminution of this peak hence suggests the lengthening of [-Cu(1)-O(1)-]$_n$ chains with overall oxygen content kept unchanged.

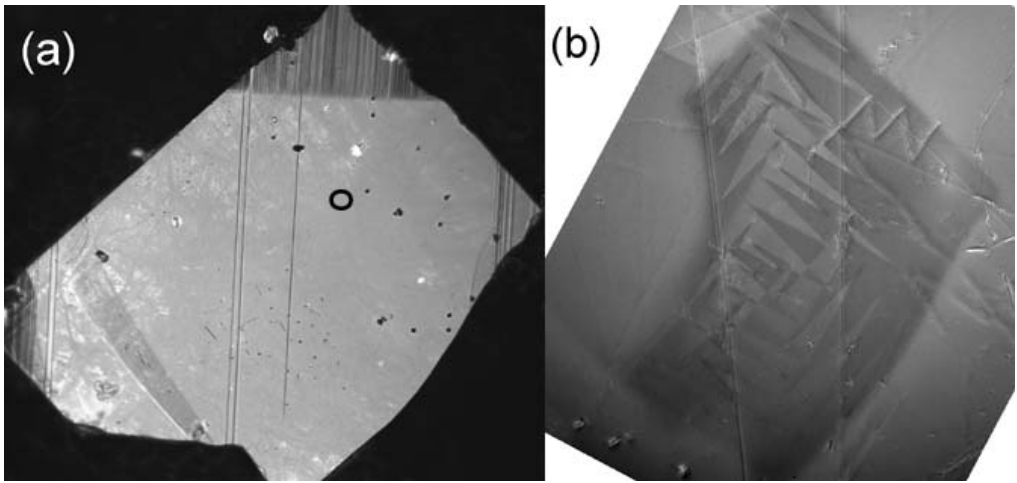

Fig1

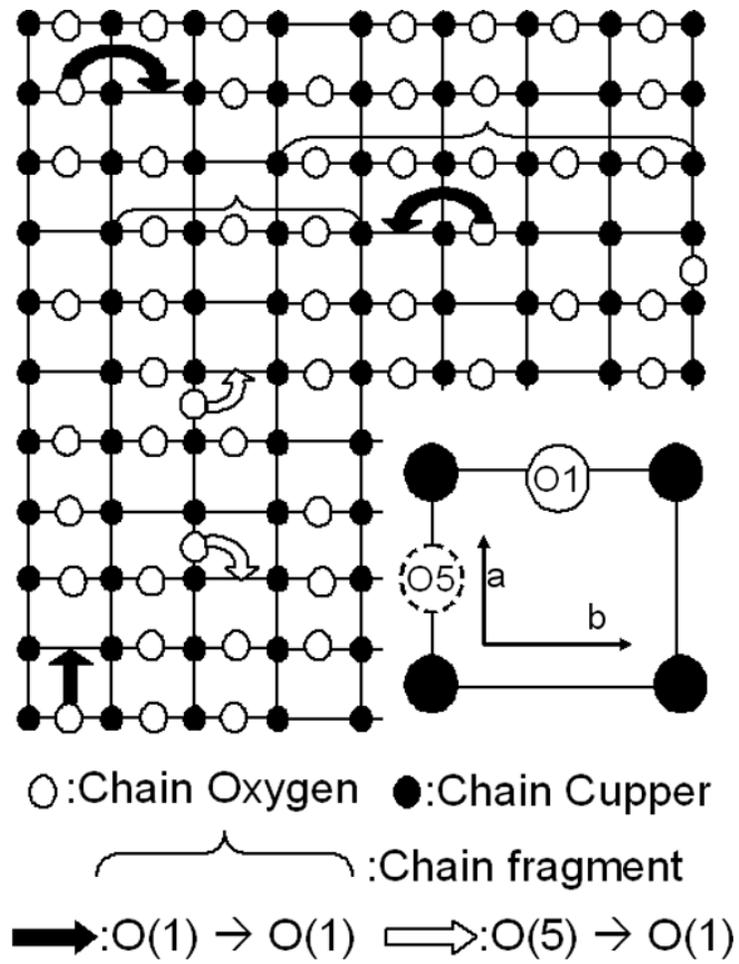

○:Chain Oxygen  ●:Chain Cupper

⌢:Chain fragment

➡:O(1) → O(1)   ⇨:O(5) → O(1)

Fig.2

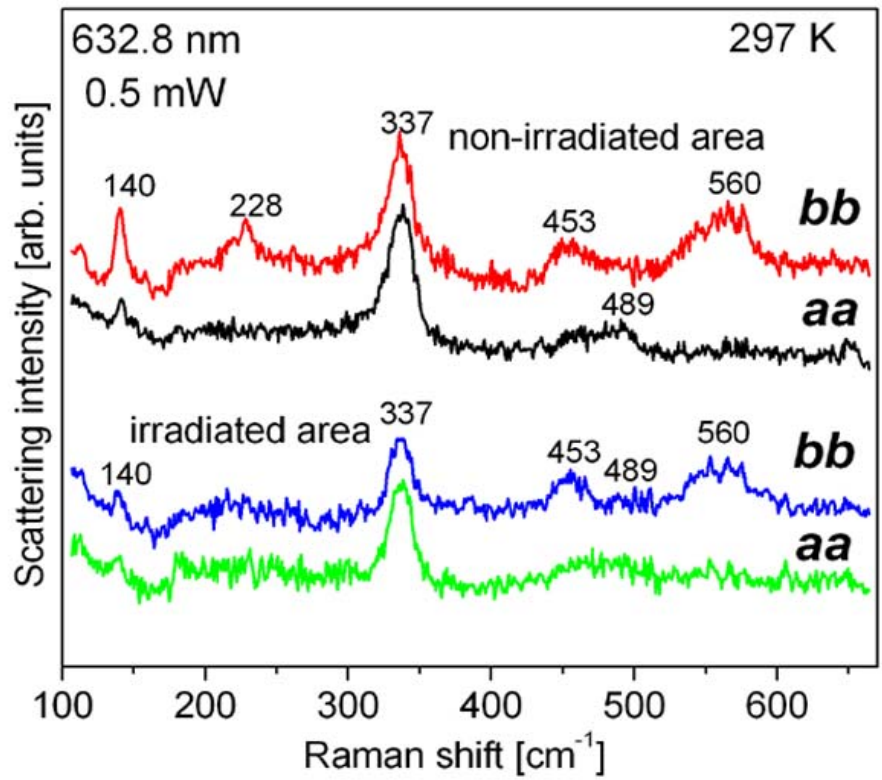

Fig. 3